\documentstyle[preprint,aps]{revtex} 

\newcommand{\lsim}{\raisebox{-0.5mm}
{$\stackrel{<}{\scriptstyle{\sim}}$}}
\newcommand{\gsim}{\raisebox{-0.5mm}
{$\stackrel{>}{\scriptstyle{\sim}}$}}

\newcommand{\ppp}[1]{%
        \setbox0=\hbox{#1}%
        \kern-.02em\copy0\kern-\wd0
        \kern+.04em\copy0\kern-\wd0
        \kern-.02em\raise.0217em\box0}
\newcommand{\vek}[1]{
        \mathchoice{\mbox{\boldmath$#1$}}%
        {\mbox{\boldmath$#1$}}%
        {\ppp{$\scriptstyle#1$}}%
        {\ppp{$\scriptscriptstyle#1$}}}

\input{epsf}

\begin{document} 
\thispagestyle{empty}

\preprint{
\vbox{
\hbox{TUM-T39-96-30}
}}

\title{Polarized deuteron structure functions at small $x$ \thanks{
Work supported in part by BMBF}}
\author{J.~Edelmann, G.~Piller and  W.~Weise}

\address{Physik Department, Institut f\"{u}r Theoretische Physik, Technische
Universit\"{a}t M\"{u}nchen, D-85747 Garching, Germany} 
\maketitle

\bigskip
\bigskip
\bigskip

\begin{abstract} 
We investigate shadowing corrections to the polarized deuteron 
structure functions $g_1^d$ and $b_1$. 
In the kinematic domain of current fixed target 
ex\-periments we observe that shadowing effects in  
$g_1^d$ are approximately twice as large as for 
the unpolarized structure function $F_2^d$.
Furthermore, we find that $b_1$ is surprisingly large at 
$x<0.1$ and receives dominant contributions from coherent 
double scattering.
\end{abstract}

\vspace*{1.cm}
\bigskip
\bigskip
{\sl \centerline {To be published in Z. Phys. A.}}

\newpage

In recent unpolarized lepton-nucleus scattering experiments at 
CERN (NMC) and FNAL (E665) \cite{data}  
nuclear shadowing at small values of the Bjorken scaling variable $x<0.1$
has been established as a leading twist effect.
It is driven by the diffractive excitation of the (virtual) photon 
into hadronic states which interact coherently with 
several nucleons in the target nucleus. 

Considering the growing interest in spin-dependent 
structure functions, a study  of shadowing 
effects in polarized deep-inelastic scattering 
is urgently needed.
In particular, the extraction of the neutron spin structure 
function $g_1^n$ from deuteron and $^3He$ data requires 
a detailed knowledge of nuclear effects in the small $x<0.1$ domain. 
Furthermore, planned experimental investigations of the 
yet unmeasured deuteron structure function $b_1$ \cite{HERMES}
call for an analysis of its small $x$ behavior which 
is driven, as we will show, by coherent double scattering contributions.
This short note summarizes our first results. 
A more complete and extended exposition of the material 
and the formalism is in preparation \cite{Edel96}.

In the following we focus on  the deuteron structure functions 
$F_{1,2}^d$, $g_1^d$ and $b_1$ at small 
values of the Bjorken scaling variable $x<0.1$.
Based on the optical theorem which connects forward virtual Compton 
scattering  and deep-inelastic scattering, 
these structure functions  
can be expressed in terms of  (virtual) photon-deuteron helicity 
amplitudes ${\cal A}^d_{+h}$, where ``$+$'' denotes the helicity of 
the transversely polarized photon and $h=0,+,-$ labels the deuteron 
helicity \cite{b1}:
\begin{eqnarray}
F_1^d &\sim& \frac{1}{3} Im \left(
{\cal A}^d_{+-}+{\cal A}^d_{++}+{\cal A}^d_{+0}
\right),\\
g_1^d &\sim& \frac{1}{2} Im \left(
{\cal A}^d_{+-}-{\cal A}^d_{++}
\right),\\
b_1 &\sim& \frac{1}{2} Im \left(
2 {\cal A}^d_{+0}-{\cal A}^d_{++}-{\cal A}^d_{+-}
\right).
\end{eqnarray}
The deuteron helicity amplitudes  
can be split into contributions from the incoherent scattering 
off  either the proton or neutron, and a term which accounts for the 
coherent scattering from both nucleons:
\begin{equation}
 {\cal A}^d_{+h} = {\cal A}^{p}_{+h}+{\cal A}^{n}_{+h}+
\delta {\cal A}_{+h}.
\end{equation}
Since nuclear effects from binding and Fermi-motion are relevant only  
at moderate and large $x\,\gsim \,0.2$ (e.g. see references in  \cite{binding}) 
we neglect them in the following. 
Then the single scattering amplitudes 
${\cal A}^{p}_{+h}$ and $ {\cal A}^{n}_{+h}$ 
are directly related to the free proton and neutron structure 
functions, respectively. 
Note that in this approximation single scattering yields no contribution 
to $b_1$.  
The double scattering amplitude $\delta {\cal A}_{+h}$ is responsible 
for shadowing corrections in $F_{1,2}^d$ and $g_1^d$ and dominates 
$b_1$ at small $x$.
Treating the deuteron target as a non-relativistic bound state,  
described by the helicity dependent wave function 
$\psi_{h}(\vek r)$ we obtain:
\begin{eqnarray} \label{eq:A2}
{\delta \cal A}_{+h}&\sim&  \sum_X
\int d^3r \int\frac{d^3k}{(2\pi)^3}\;
\psi_{h}^{\dagger}(\vek{r}) \;{\cal T}(\gamma^* p \rightarrow Xp)\, 
 \nonumber \\
&\times&\;\frac{e^{-i\vek{k}\cdot\vek{r}}}
{\nu^2 - \vek k_{\perp}^2 - (q_z-k_z)^2 - M_X^2 + i \epsilon}\;{\cal T}(X n \rightarrow \gamma^* n) \;\psi_{h}(\vek{r})\;+\;[p\leftrightarrow n],
\end{eqnarray}
where  $q^{\mu} = (\nu,{\bf 0}_{\perp},q_z)$ 
is  the four-momentum of the virtual photon. 
The sum is taken over all diffractively excited hadronic 
intermediate states with momentum $q-k$ and invariant mass $M_X$.  
The amplitudes 
\begin{eqnarray}
{\cal T}(\gamma^* p \rightarrow Xp) &=& 
P^p_{\uparrow}t^{\gamma^*p\to Xp}_{+\uparrow}+
P^p_{\downarrow}t^{\gamma^*p\to Xp}_{+\downarrow},
\\
{\cal T}(X n \rightarrow \gamma^* n) &=&
P^n_{\uparrow}t^{Xn\to \gamma^*n}_{+\uparrow}+
P^n_{\downarrow}t^{Xn\to \gamma^*n}_{+\downarrow}
\end{eqnarray}
stand for  proper combinations of 
projection operators $P^{p(n)}_{\uparrow(\downarrow)}$ onto 
proton (neutron) states with helicity $+1/2$ $(-1/2)$, and 
$t^{\gamma^*p\to Xp}_{+\uparrow}$ etc.,
the corresponding  photon-nucleon helicity amplitudes 
for the diffractive production of the state $X$.

Coherent double scattering contributions to the deu\-teron 
structure functions are determined by the  
imaginary part of $\delta {\cal A}_{+h}$ 
which is dominated by the diffractive production and 
re-scattering of hadronic states in  forward direction. 
The result for $\delta {\cal A}_{+h}$ can then be expressed 
in terms of helicity-dependent 
diffractive photon-nucleon forward amplitudes 
(which we assume to be purely imaginary), 
combined with the longitudinal deu\-teron form factor
\begin{equation}
{\cal F}_{h}(1/\lambda)=\int_{-\infty}^\infty
dz\;|\psi_{h}({\bf 0}_{\perp},z)|^2 \cos({z}/{\lambda}),
\end{equation}
for each deuteron helicity $h=0,+,-$. 
These form factors are functions of the inverse 
propagation length $\lambda^{-1}$;
a hadronic fluctuation of mass $M_X$ can contribute to coherent 
double scattering only if its propagation length 
$\lambda = 2\nu/(Q^2 + M_X^2)$ exceeds the deuteron size: 
$\lambda \,\gsim \,\langle r^2 \rangle_d ^{1/2} \approx 4\,fm$.  

\bigskip
\noindent{\bf Unpolarized structure function}

Neglecting any spin- and isospin-dependence of the diffractive 
photon-nucleon amplitudes yields the standard result 
for shadowing in the  unpolarized structure function  
$F_1^d=F_1^p+F_1^n+\delta F_1$ (e.g. see references in \cite{binding}):
\begin{equation} \label{eq:delta_F1}
\delta F_1(x,Q^2)=-\frac{Q^2}{x\pi \alpha}
\int dM_X^2 
\left.\frac{d^2\sigma^{\gamma^*N}}
{dM_X^2 dt}\right|_{t\approx 0}{\cal F}(1/{\lambda}),
\end{equation}
where $\alpha = 1/137$. Here 
${\cal F} = ({\cal F}_+ + {\cal F}_- + {\cal F}_0)/3$ 
is the helicity averaged longitudinal deuteron form factor, 
and $d^2\sigma^{\gamma^*N}/dM_X^2 dt$ is the 
unpolarized forward cross section 
for the diffractive production of hadronic 
states $X$ from nucleons. 
Corrections to eq.(\ref{eq:delta_F1}) are discussed in \cite{Edel96} (see also \cite{HK93}).

\bigskip
\noindent{\bf Polarized structure function ${\bf g_1^d}$}

In $g_1^d=(1-\frac{3}{2}\omega_D)(g_1^p+g_1^n)+\delta g_1$ 
the single scattering contribution is modified by 
the $D$-state probability $\omega_D$. 
Assuming isospin invariance of the unpolarized diffractive 
photon-nucleon amplitudes leads to:
\begin{eqnarray} \label{eq:delta_g1}
\delta g_1(x,Q^2)&=&-\frac{Q^2}{4x\pi\alpha}\int dM_X^2
\left[
\left.
\frac{d^2\sigma^{\gamma^*p}_{+\downarrow}}
{dM_X^2 dt}\right|_{t\approx 0}-\left.\frac{d^2\sigma^{\gamma^*p}_{+\uparrow}}
{dM_X^2 dt}\right|_{t\approx 0} 
\right] 
{\cal F}_+(1/\lambda) + 
[p \leftrightarrow n].
\end{eqnarray}
A direct comparison of the double scattering helicity amplitudes  
yields the following upper limit: 
\begin{equation} \label{eq:est}
\frac{\left|\delta g_1\right|}{F_1^N}
\le \frac{3\left|\delta F_1\right|}{2 F_1^N} 
\approx \frac{3\left|\delta F_2\right|}{2 F_2^N}, 
\end{equation}
where $F^N_{1,2} = (F^p_{1,2} + F^n_{1,2})/2$. 
Data on ${\delta F_2}/2 F_2^N$  
are available from E665 \cite{data} (see Fig.1a). 
We conclude  from eq.(\ref{eq:est}) that, 
for  recent data analyses 
\cite{g1d} on the neutron structure function $g_1^n$,
uncertainties due to the 
shadowing correction $\delta g_1$ are within the experimental 
errors. 
Note that the upper bound (\ref{eq:est}) is not helpful  
at very small values $x<0.01$.

In a laboratory frame description at $x<0.1$, the virtual 
photon fluctuates into a hadronic state 
which then interacts with one or several nucleons inside the 
nuclear target.
In the kinematic range of currently available experimental data 
on shadowing in unpolarized lepton scattering, 
it has turned out to be a good approximation 
to consider the interaction of only one effective hadronic 
state with invariant mass $M_X^2 \sim Q^2$ and a coherence 
length $\lambda \sim 1/2Mx$. 
This ``one-state'' approximation has been recently applied to   
shadowing in $g_1^{^3He}$ \cite{FrGuSt96}. 
For deuterium it yields: 
\begin{equation}\label{Verhaltnis}
\frac{\delta g_1}{g_1^N} = 
{\cal R}_{g_1} \frac{\delta F_2}{F_2^N},
\end{equation}
with ${\cal R}_{g_1} = {\cal F}_+(2Mx)/{\cal F}(2Mx)$.
At $x\,\lsim \,0.01$ we obtain for realistic deuteron wave functions 
${\cal R}_{g_1}=2.7$ (Paris potential \cite{LaLoRi80}) 
and ${\cal R}_{g_1}=2.4$ (Bonn one-boson exchange potential \cite{Bonn}).
In Fig.1b we show the shadowing correction $\delta g_1/2 g_1^N$ 
using recent data on $\delta F_2/2 F_2^N$ from the E665 collaboration 
\cite{data}.
It should be noted that for decreasing values of $x$ 
the ex\-perimental data for the shadowing ratio $\delta F_2 /2 F_2^N$ 
are taken  at decreasing values of the average momentum transfer 
$\overline {Q^2}$.
Therefore our result for $\delta g_1$ shown in Fig.1b  
corresponds, strictly speaking, to the fixed target 
kinematics of E665 \cite{data} which is not 
far from the kinematics of  SMC \cite{g1d}.

In the kinematic domain of recent experiments one has 
$|g_1^N| = |g_1^n + g_1^p|/2 < 0.5$ \cite{g1d}. 
Returning to eq.(\ref{Verhaltnis}) one then observes that shadowing  amounts 
to less then $5\%$ of the experimental 
error on $g_1^n$ for the  SMC analysis \cite{data}. 

\bigskip
\bigskip
\bigskip
\noindent
{\bf Polarized structure function ${\bf b_1}$}

At small values of $x<0.1$ coherent double scattering 
dominates $b_1$ and leads to:
\begin{eqnarray} \label{eq:delta_b1}
b_1(x,Q^2)&=&
\frac{Q^2}{x\pi \alpha}
\int dM_X^2
\left.
\frac{d^2\sigma^{\gamma^*N}}{dM_X^2 dt}
\right|_{t\approx 0}\left(
{\cal F}_{+}(1/\lambda)-
{\cal F}_{0}(1/\lambda)
\right),
\end{eqnarray}
where binding effects are small \cite{b1}.
Here we have again neglected any spin- and isospin-dependence of the 
diffractive photon-nucleon amplitudes \cite{Edel96}. 
At small $x\,\lsim \,0.01$ a good  approximation to 
eqs.(\ref{eq:delta_F1},\ref{eq:delta_g1},\ref{eq:delta_b1})
is obtained by setting ${\cal F}_h(1/\lambda)\approx {\cal F}_h(0)$ 
for hadronic states with $\lambda > \langle r^2\rangle_d^{1/2}$, 
while  ${\cal F}_h(1/\lambda)=0$ otherwise \cite{PNW96}. 
We then find for the double scattering contribution to $b_1$:
\begin{equation} \label{eq:b1_double}
\frac{b_1}{F_1^N}=
{\cal R}_{b_1} \,\frac{\delta F_2}{F_2^N}, 
\end{equation}
with ${\cal R}_{b_1}= ({\cal{F}}_{0}(0)-{\cal F}_{+}(0))/{\cal F}(0)$.
Note that ${\cal R}_{b_1}=0$ if the $D$-state admixture in the deuteron 
is neglected. 
We obtain ${\cal R}_{b_1}=-1.03$ (Paris potential \cite{LaLoRi80}) 
and ${\cal R}_{b_1}=-0.58$ (Bonn one-boson exchange potential \cite{Bonn}).
With recent data for $F_2^N$ \cite{E665FN} combined with the 
Callan-Gross relation, and the measured 
$\delta F_2/2F_2^N$ \cite{data} 
we find a large contribution to $b_1$ 
from coherent double scattering as shown in Fig.2. 
This is a remarkable result. As we found after the 
present calculations \cite{Edel96} were finished, a similar conclusion has
recently  been reached in ref.\cite{NikSch96}.

In summary, we have presented first results for  
shadowing corrections in polarized deep-inelastic 
scattering from deuterium.
In the kinematic regime of current fixed target 
ex\-periments,  
shadowing in  $g_1^d$ is found to be approximately twice 
as large as for the unpolarized structure function $F_2^d$.
Nevertheless it plays a minor role for the extraction 
of the neutron structure function $g_1^n$.
Furthermore we find dominant contributions to the deuteron 
structure function $b_1$ at $x<0.1$ from 
coherent double scattering involving the deuteron $D$-state.


\begin{figure}
\centerline{\epsfxsize=0.75\hsize \epsffile{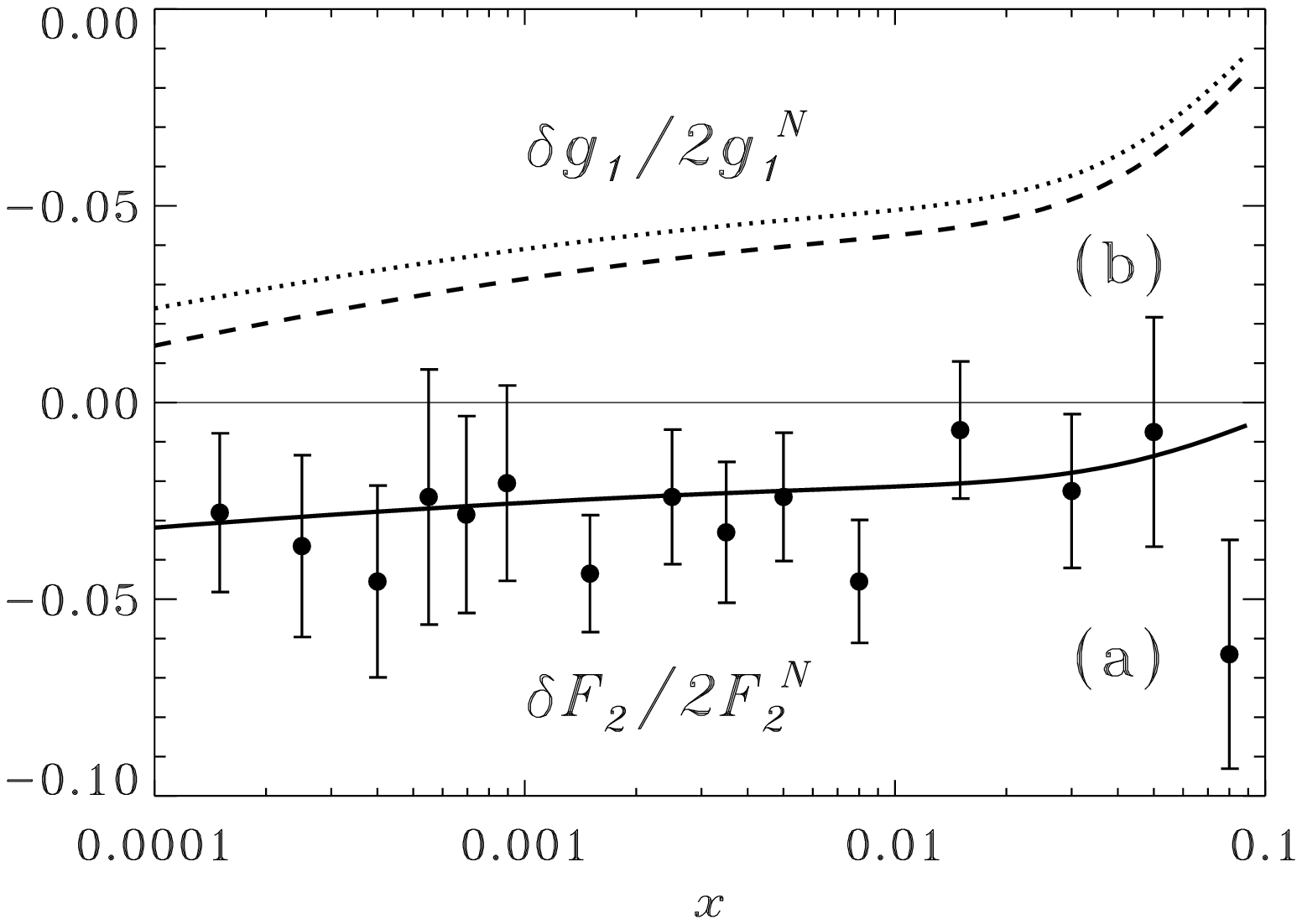}}
\caption[]{(a) shadowing correction $\delta F_2/2F_2^N$, 
data from E665 \cite{data}. 
The full line represents a parametrization of the data 
used in  (\ref{Verhaltnis}) and (\ref{eq:b1_double}).
(b) shadowing correction $\delta g_1/2 g_1^N$ from 
(\ref{Verhaltnis}). 
The dashed and dotted curves  correspond to the Paris 
\cite{LaLoRi80} and Bonn \cite{Bonn} potential respectively.}
\end{figure}

\begin{figure}
\centerline{\epsfxsize=0.75\hsize \epsffile{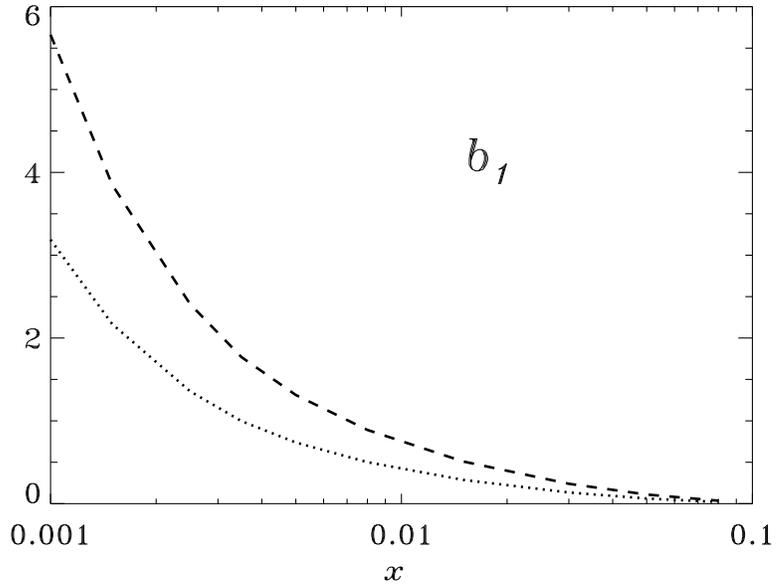}}
\caption[]{Double scattering contribution to $b_1$ 
from (\ref{eq:b1_double}).
The dashed and dotted curves  correspond to the Paris 
\cite{LaLoRi80} and Bonn \cite{Bonn} potential respectively.}
\end{figure}

\end{document}